\documentclass[11pt,twoside]{article}
\usepackage{./asp2014}
\usepackage{color}

\aspSuppressVolSlug
\resetcounters

\begin{document}

\title{Modeling the Spectra of Dense Hydrogen Plasmas: Beyond
  Occupation Probability}
\author{T. A. Gomez,$^{1,2}$, M. H. Montgomery$^1$, T. Nagayama$^2$,
  D. P. Kilcrease$^3$, and D.~E. Winget$^1$}
\affil{$^1$Department of Astronomy, University of Texas, Austin, Texas
  78712,
  USA; \email{gomezt@astro.as.utexas.edu}}
\affil{$^2$Sandia National Laboratories, Albuquerque, New Mexico 87185, USA}
\affil{$^3$Los Alamos National Laboratories, Los Alamos, New Mexico
  87545, USA}

\paperauthor{Sample~Author1}{Author1Email@email.edu}{ORCID_Or_Blank}{Author1 Institution}{Author1 Department}{City}{State/Province}{Postal Code}{Country}
\paperauthor{Sample~Author2}{Author2Email@email.edu}{ORCID_Or_Blank}{Author2 Institution}{Author2 Department}{City}{State/Province}{Postal Code}{Country}
\paperauthor{Sample~Author3}{Author3Email@email.edu}{ORCID_Or_Blank}{Author3 Institution}{Author3 Department}{City}{State/Province}{Postal Code}{Country}

\begin{abstract}
  Accurately measuring the masses of white dwarf stars is crucial in
  many astrophysical contexts (e.g., asteroseismology and
  cosmochronology). These masses are most commonly determined by
  fitting a model atmosphere to an observed spectrum; this is known as
  the spectroscopic method. However, for cases in which more than one
  method may be employed, there are well known discrepancies between
  masses determined by the spectroscopic method and those determined
  by astrometric, dynamical, and/or gravitational-redshift
  methods. In an effort to resolve these discrepancies, we are
  developing a new model of hydrogen in a dense plasma that is a
  significant departure from previous models.  Experiments at Sandia
  National Laboratories are currently underway to validate these new
  models, and we have begun modifications to incorporate these models
  into stellar-atmosphere codes.
\end{abstract}

\section{Astrophysical Context}

White dwarf stars are of critical importance in many contexts, from
defining the initial-final mass relation of low and
intermediate mass stars \citep[e.g.,][]{Catalan08,Williams09} to
acting as independent chronometers for different components of the
Galaxy \citep[e.g.,][]{Winget87,Garcia-Berro10}.

For all of these investigations we require accurate estimates of the
masses and temperatures of these stars. By far the most widely used
technique, termed the ``spectroscopic method'', involves fitting a
model atmosphere to an observed spectrum of a star, and varying the
$T_{\rm eff}$ and $\log\,g$ values to match the width and intensities
of the spectral lines \citep{Barstow05}, and optionally the continuum as well. The set
of best-fit parameters are then identified as estimates of the
temperature and surface gravity of the white dwarf.

\section{Sources of Uncertainty}

Problems in applying this method can arise from many sources of
uncertainty: surface composition, the physics of convection, atomic
physics, magnetic fields, etc. Recently, progress has been made on the
effect of convection on the parameters derived from spectroscopic fits
of white dwarfs \citep[e.g.,][]{Tremblay11,Tremblay13}, but the other sources
of uncertainty remain.

\begin{figure}[!t]
\centering{
\includegraphics[width=0.9\textwidth,angle=0]{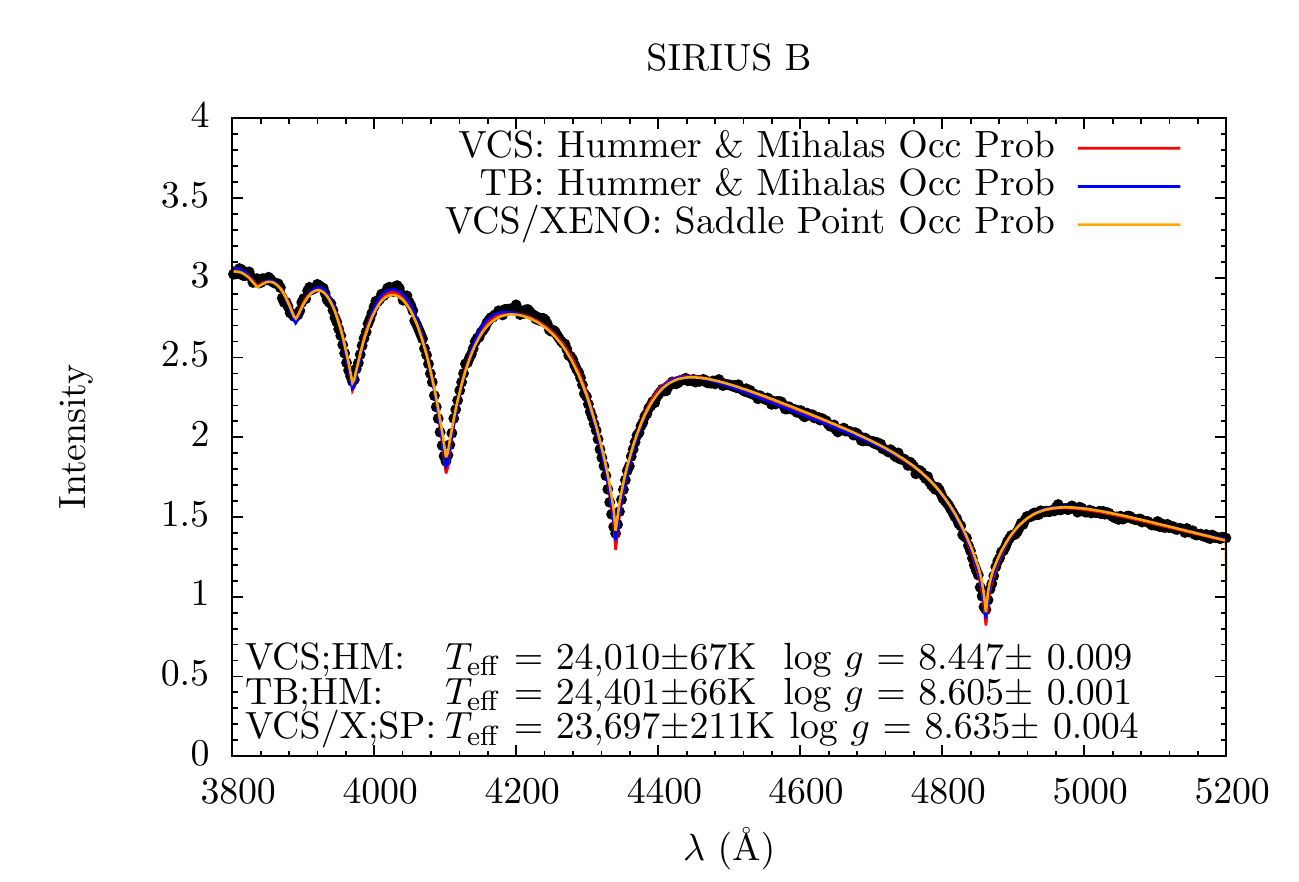}
}
\caption{Three fits to a spectrum of Sirius B (HST STIS). The differences in
  derived parameters for each fit are due solely to the different
  assumed atomic physics of line profiles and the occupation
  probability of atomic states.}
\vspace*{-0.5em}
\label{sirius}
\end{figure}

As an illustration of this, in Figure~\ref{sirius} we show three
different fits to a spectrum of Sirius~B \citep{Barstow05}. All the
fits reproduce the spectrum quite well but derive different estimates
for the $T_{\rm eff}$ and $\log\,g$ of the star. The same
model-atmosphere code Tlusty \citep{Hubeny94,Hubeny95} was used to
compute these fits; the only differences arose from the treatment of
the atomic physics of line profiles and the occupation probability
(OP) of atomic states.  Tlusty includes OP put forth by
\citet{Dappen87} and \citet{Hummer88}.  The \citet{Hummer88} OP
defines the ionization threshold by the \citet{Inglis39} criteria.
However, there are other studies that show different ionization
criteria \citep{Luc-Koenig80}.  We therefore calculated model spectra
with a different ionization criteria (defined by the saddle point in
the potential) to compare the possible uncertainties in the $\log\,g$
(yellow line in Fig \ref{sirius}).

To resolve this discrepancy we would hope to appeal to independent
techniques for determining Sirius B's parameters. Unfortunately,
astrometric and gravitational-redshift techniques disagree with each
other, and with the spectroscopic method \citep[e.g.,][]{Barstow15}.
Thus, the time seems ripe to re-visit the physics of line
formation in white dwarf atmospheres.

\section{New Line-Profile Calculations}

For an isolated atom, the atomic structure can often be calculated
quite accurately.  This is certainly the case for the hydrogen atom,
since both the Schr\"odinger and Dirac equations have exact
solutions. On the other hand, if the atom is embedded in a plasma,
then it inherently becomes a many-body problem, and solutions are
possible only within a set of approximations and methods.

\begin{figure}[!t]
\includegraphics[width=1.0\textwidth,angle=0]{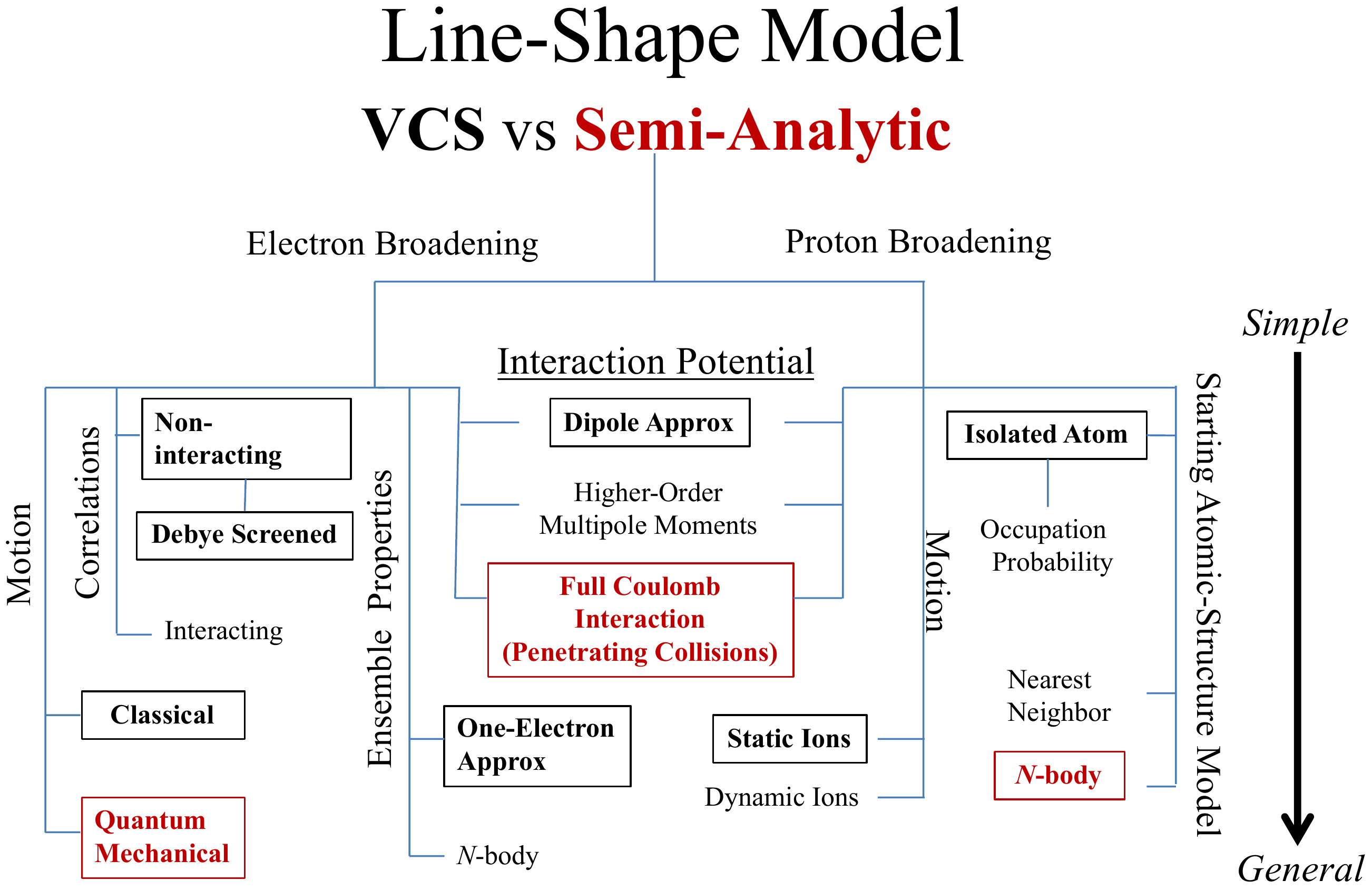}
\caption{The ``Approximation Tree'' for the theory of line-profile
  modeling. The black,and bold-faced text boxes show the
  approximations used in the original VCS theory \citep{Vidal73}, and
  the corresponding red, bold-faced items show the approximations used
  in our new semi-analytic approach.}
\label{tree}
\end{figure}

In Figure~\ref{tree}, we depict the various approximations and
approaches commonly used in line-profile calculations. The black,
bold-faced text boxes correspond to the approximations adopted by
\citet{Vidal73} and are commonly termed ``VCS theory'', while the red,
bold-faced text boxes show the set of approximations used in our new
``semi-analytic'' theory.  We note that the recent improved
line-profile calculations of \citet{Tremblay09} are essentially VCS
theory plus the \citet{Hummer88} and \citet{Seaton90} formalism for
treating continuum lowering and level dissolution.  \citet{Hummer88}
and \citet{Seaton90} define an ionization criteria for a given energy
level, then perform a statistical average over plasma electric fields
to determine the probability that the state is still ``occupied'';
this is known as occupation probability.

We have begun a program to re-examine the approximations shown in
Figure~\ref{tree} with an eye to which ones can be improved. To date,
we have employed two approaches to this problem: 1) a simulation-based
($N$-body) approach, and 2) a semi-analytic approach. The simulation
approach treats the perturbing influences of the protons and electrons
classically and then solves the time-dependent Schr\"odinger equation
numerically to obtain the broadened profiles. The semi-analytic
approach treats the perturbing electrons quantum mechanically
\citep[][Gomez et al., in prep]{Baranger58,Fano63} and then solves the
time-\emph{independent} Schr\"odinger equation for the hydrogen atom
in the presence of several nearby protons.  Both methods are able to
explore different approximations and physical effects. For instance,
we have used the simulation approach to explore ion dynamics 
\citep[not discussed here; see ][for more details]{2014A...2..299} 
and the importance of higher-order multipoles \citep[e.g.,][]{Gomez16} of the
interaction potential. We will use the semi-analytic approach to
examine the importance of a quantum treatment of electrons and
an ab initio treatment of continuum lowering \citep{Crowley14}.

\begin{figure}[!t]
\centering{
  \includegraphics[width=0.80\textwidth,angle=0]{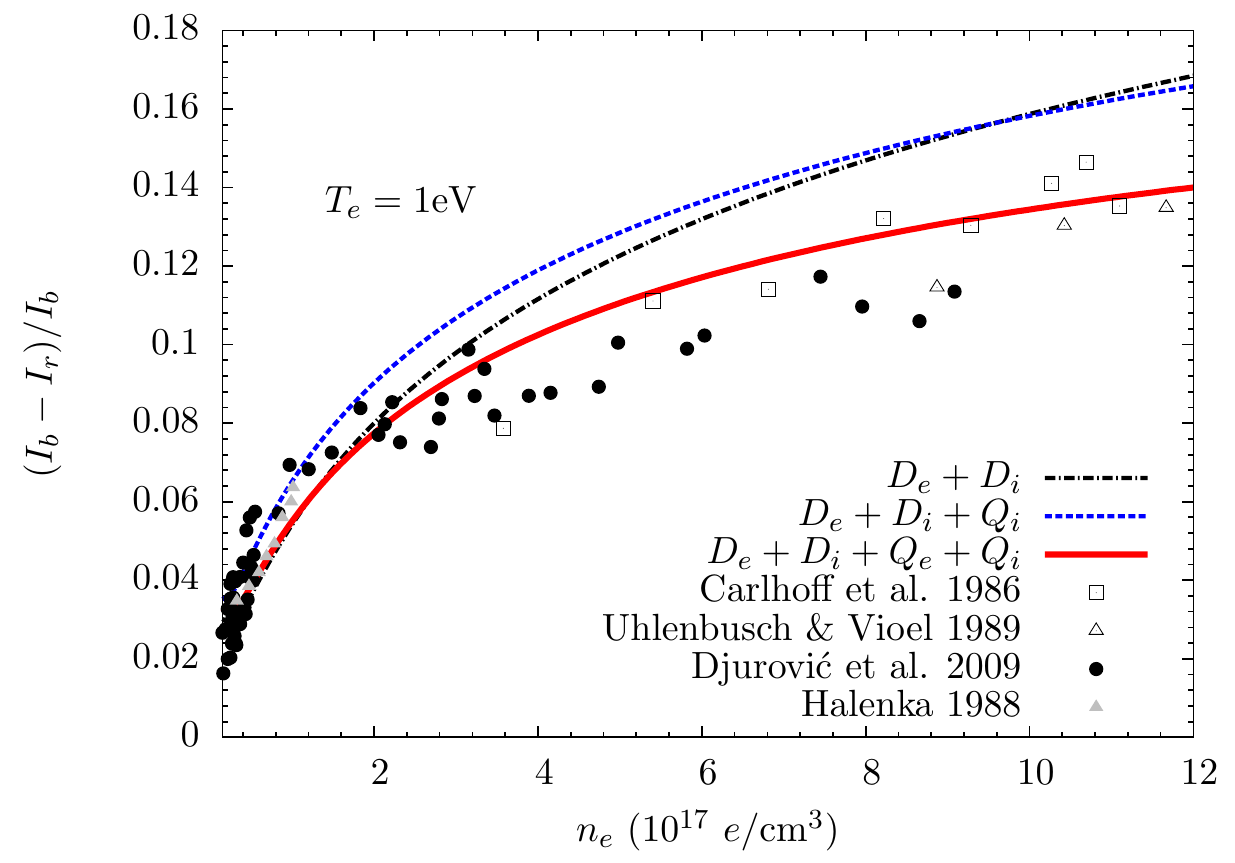}
}
\caption{ The red/blue asymmetry of H$\beta$ as a function of density
  \citep[taken from][]{Gomez16}. In the legend, $D_e$ and $Q_e$
  indicate the dipole and quadrupole terms due to electrons,
  respectively, and similarly for the ions ($D_i$ and $Q_i$).  As
  indicated, the data (different style points) are taken from
  \citet{Carlhoff86}, \citet{Djurovic09}, and \citet{Uhlenbusch89}.  }
\label{asym}
\end{figure}

\subsection{Higher-order Multipole Moments}
We have recently published results that improve the treatment of the
electrostatic interaction potential.  In \citet{Gomez16}, we use a
simulation method to examine the importance of the higher-order terms
in the multipole expansion of the interaction potential and the
validity of the dipole approximation. While most calculations include
only the dipole term, corresponding to a spatially constant electric
field, we explored the effect which including higher-order gradients
in the field (the quadrupole, octupole, and sedecapole terms) had on
the detailed line-profile shape. For electron densities up to
$10^{18}\,$e/cc, we found that it was necessary to retain up to the
quadrupole terms for both electrons and ions in the multipole
expansion to correctly infer the density to within a few percent. In
addition, these higher-order terms allowed us to match the height
asymmetry seen in the ``red'' and ``blue'' central peaks of H$\beta$
(solid line in Figure~\ref{asym}).  This is the first time that a
theoretical calculation has been able to match this asymmetry for
densities of $\sim 10^{17}$--$10^{18}\,$e/cc.  While this feature is
not usually seen in white dwarf spectra, it can be used as a
density diagnostic for laboratory plasmas \citep{Djurovic09}.

\subsection{$\pmb{N}$-body Treatment of Protons \& Quantum Electrons}
\label{sa}

The red text boxes in Figure~\ref{tree} show the approximations used
in our semi-analytic approach. It differs from VCS (or TB) in that it:
1) has a \emph{quantum} treatment of the perturbing electrons, 2) uses
a full Coulomb interaction of the perturbing electrons with the
radiating-atom wave function, including penetrating collisions, and 3)
computes an exact numerical solution of the Schr\"odinger equation in
the $N$-body potential of the nearby ions.

At high densities, the electric fields of nearby protons cause the
energy levels of the radiating atom to begin to cross; this is termed
the ``Inglis-Teller limit'' \citep{Inglis39}.  The current reasoning
\citep{Hummer88} is that when energy levels cross, then the state is
considered dissolved into the continuum.  Preliminary results for the
proton-dependent energy-level structure show that discrete states
exist beyond the Inglis-Teller limit, but, due to the averaging of the
different proton configurations and the electron broadening, the
spectrum appears featureless above this threshold. This approach
produces a different spectral shape for lines near the continuum
compared to the standard treatment; this could result in
systematically different inferred masses for white dwarf stars.

\section{Discussion}

\begin{figure}[!t]
\centering{
\includegraphics[width=0.8\textwidth,angle=0]{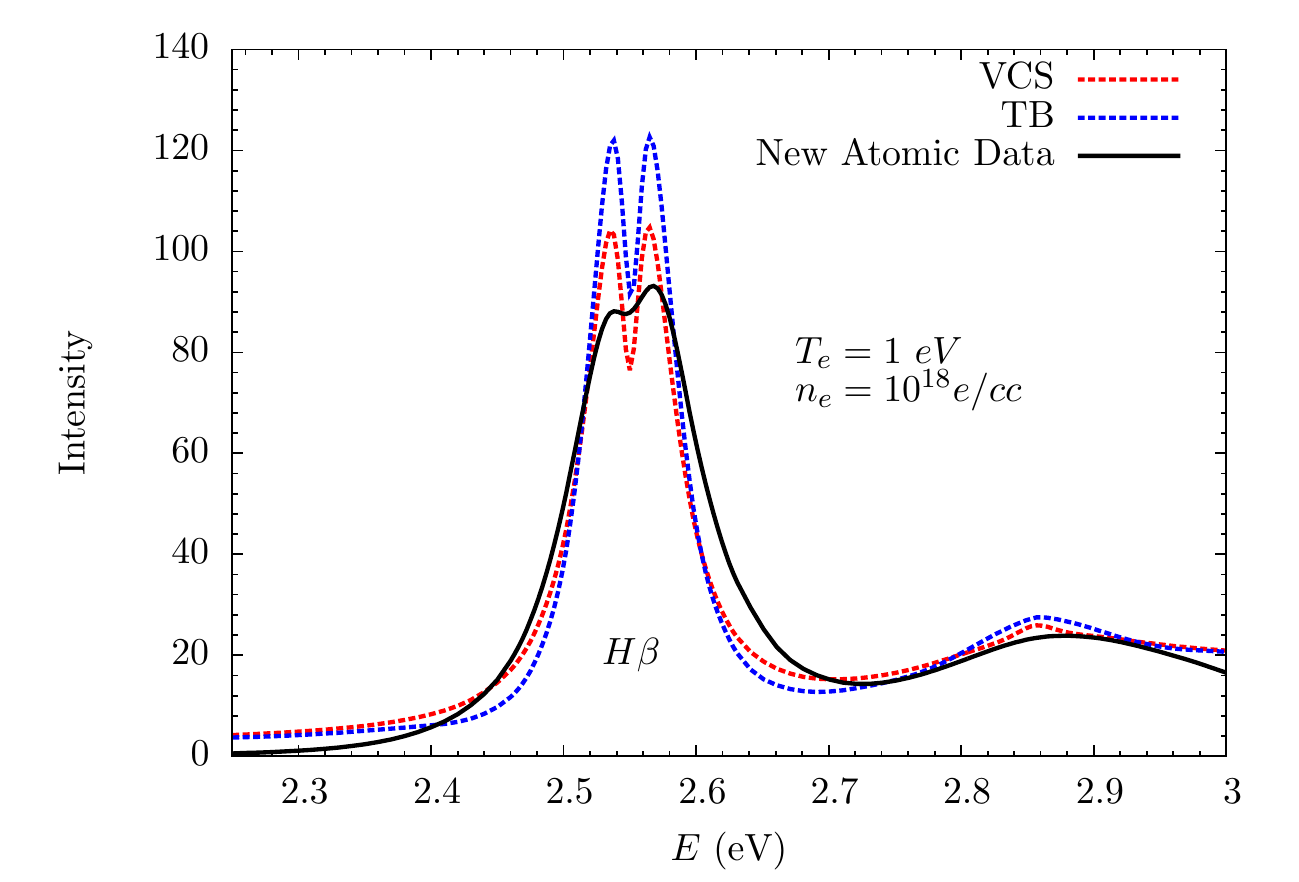}
}
\caption{A preliminary comparison of VCS, TB, and our new
  semi-analytic profiles for a temperature of $1\,\rm eV$ and a
  density of $10^{18}\,$e/cc.  }
\label{hb2}
\end{figure}

We do not yet have results for the calculations using the approach
described in section~\ref{sa}. As an illustration, however, in
Figure~\ref{hb2} we show a comparison of a preliminary calculation
using our semi-analytic model with the results of VCS and TB models.

Our new $N$-body atomic-structure calculation reveals many discrete
states between 2.9 and 3.0~eV.  However, when an average is taken over
different free-proton configurations, these transitions are no longer
discernible and the spectrum appears to be continuous. This direct
treatment of proton perturbations results in a lowered opacity redward
of the H$\beta$ line compared to the VCS and TB calculations.
Finally, we note that our quantum electron-broadening treatment
produces broader H$\beta$ and H$\gamma$ profiles for these conditions
($T_e = 1\,{\rm eV}$, $n_e=10^{18} \,\rm e/cc$).

This ab initio calculation illustrates the potential power of this approach: we
are able to solve for eigenstates spanning a broad range of
conditions, from tightly bound and minimally perturbed to beyond the
Inglis-Teller limit, all without recourse to an ad-hoc occupation-probability
treatment. Besides providing a more self-consistent treatment of line
profiles, this approach could lead to a parameterization of the
``effective occupation probability'' of different atomic states that
could be incorporated into model-atmosphere and other atomic
codes.

\acknowledgements 
T.A.G. acknowledges support from the National Science Foundation
Graduate Research Fellowship under grant DGE-1110007. M.H.M. and
D.E.W. acknowledge support from the United States Department of
Energy under grant DE-SC0010623.

\end{document}